\documentstyle[12pt]{article}
\def\ii{\'\i}
\def\beq{\begin{equation}}
\def\endq{\end{equation}}
\def\nat{I\kern-.25em N}
\def\uno{1\kern-.24em \text{I}}
\def\RR{ I\kern-.25em R}
\def\PP{ I\kern-.25em P}
\def\lin{\hbox to 0.7truecm{\linfill}}
\def\ZZ{ Z\kern-1.0em Z \kern+.2em}
\def\ZZsub{ Z\kern-0.4em Z\kern+.1em}

\begin{document}
\begin{center}
{\bf On the existence of monotonic fronts for a class of \\
     physical problems described by the  equation $\lambda u''' + u' = f(u)$}
\end{center}
\vspace{.1cm}
\begin{center}
R.\ D.\ Benguria and M.\ C.\ Depassier\\
        Facultad de F\ii sica\\
        P. Universidad Cat\'olica de Chile\\
               Casilla 306, Santiago 22, Chile
\end{center}

\date\today
\begin{abstract}
We obtain an upper bound on the value of $\lambda$ for which monotonic front
solutions of the equation $\lambda u''' + u' = f(u)$ with $\lambda > 0$
 may exist.
\end{abstract}

\bigskip
\bigskip
\bigskip
\bigskip

\noindent
AMS classification scheme numbers: 34 {} C {} xx, 58 {} F xx.

\newpage

\section{Introduction}
In a variety of physical phenomena the structure of fronts is described by a
third order differential equation of the form
\beq
\lambda w''' + w' = f(w), \qquad \lambda > 0, \quad x \in \RR,
\endq
where primes denote derivatives with respect to $x$,
f is a positive and continuous function for $w \in (-1,1)$ and such that
$f(-1) = f(1) = 0$.  For example, equation (1.1) with
 $f(w) = \cos ({{\pi w}\over{2}})$ and
$\lambda$ small arises in the geometric model of crystal growth
\cite{BKKL84,BGKL84}. A more
complicated version of the geometric model of crystal growth is  given by
equation (1.1) with $f(w) = \cos ({{\pi w}\over{2}})/(1 + \alpha \cos (2\pi
w))$,
where $0 < \alpha < 1$ represents crystalline anysotropy.
Traveling wave solutions of the Kuramoto-Sivashinsky equation
which arises in the study
of reaction diffusion systems \cite{KSuzuki76}, flame propagation
\cite{Siv77}, and others,  obey  the
above equation with $f(w) = 1 - w^2$.
In this latter case $\lambda = (c/2)^2$ where $c$ is the speed of the
travelling wave.
Our goal in this article is to determine generic bounds on $\lambda$ for which
equation (1.1) has no monotonic fronts; i.e. has no solutions $w$ with $w' > 0$
and such that $\lim_{x \rightarrow - \infty} w = -1, \quad
\lim_{x \rightarrow \infty} w = 1$. For the case $f(w) = 1- w^2$ bounds of this
sort were found by Toland \cite{Toland88}.
 In fact, he proved that for $\lambda \ge 2/9$
there is no monotonic solution of (1.1) on $\RR$. Although Toland's bound is
certainly correct, for the case in question it is now known that for all
$\lambda > 0$ there is no monotonic solution of (1.1) (see e.g.
\cite{JTM92,GH91}).
For the equation describing needle crystals (i.e. for
 $f(w) = \cos ({{\pi w}\over{2}})$) it has also been shown that no monotonic
solutions exist \cite{HM89,AM90}. In spite of these negative results, there are
explicit examples
of functions $f$ for which monotonic fronts do exist. This is the case for the
modified equation of the geometric model of crystal growth (i.e. for
$f(w) = \cos ({{\pi w}\over{2}})/[1 + \alpha \cos (2\pi w)]$\,) for a discrete
set
of values of the crystal anisotropy parameter $\alpha$ \cite{KS91}.
 A simpler example for
which monotonic fronts exist is given by $f(w) = {1\over 2} (1-w^2)(1- {\lambda
\over 2} + {3\over 2} \lambda w^2)$, for which the front
 $w(x) = {{e^x-1}\over{1+e^x}} = \tanh ({x\over 2})$ is monotonic, satisfies
equation (1.1) for this $f$ and also the boundary values $\lim_{x\rightarrow
\pm
\infty} w = \pm 1$. Moreover, one can construct many other explicit examples of
$f's$ for which equation (1.1) exhibits monotonic fronts.
Here we prove a generic bound on the values of $\lambda$ for which equation
(1.1)
together with the boundary values $\lim_{x\rightarrow \pm
\infty} w = \pm 1$ does not have monotonic solutions. Our main result is the
following

\vspace{.5cm}
{\bf Theorem}
{\it If
\beq
\lambda > \frac{.228}{(\int_{-1}^{1} f(t) (1-t^2) dt)^2} \quad,
\endq
then there is no solution of
\[ \lambda w'''+ w' = f(w)
\]
satisfying $\lim_{x\rightarrow \pm
\infty} w = \pm 1$ and $w' \ge 0$ on $\RR$.
}

Several remarks are in order concerning this result. First, for the case
considered by Toland, that is for $f(w) = 1 - w^2$, our bound is slightly
better
than his ($\lambda_{Toland} = 2/9 \approx 0.222, \lambda_{here} = .201$ ),
although we know that both of these bounds are not relevant because of the non
existence results of Jones et el \cite{JTM92} ( see also \cite{GH91,Grimshaw92}
 ). Second, our bound
is not optimal, in the sense that there is no $f$ for which inequality (1.2)
is saturated (i.e. satisfied as an inequality). Third, the methods used here to
prove bounds on $\lambda$ for which there are no monotonic solutions can be
easily extended to treat more general equations. In particular
they have been used by us to determine bounds on the speed of monotonic
travelling fronts of a Kuramoto-Sivashinsky equation with
dispersion \cite{BD93}.
 We
 do not attempt to prove the existence of fronts, which requires an entirely
different approach \cite{Toland88}. The rest of the paper is organized as
follows: in Section 2 we prove the bound
and in Section 3 we apply our bound to several examples.

\section{Proof of the bound on $\lambda$}

\setcounter{equation}{0}
Here we are only concerned about  monotonic solutions $w(x)$ of equation (1)
satisfying $w(x) \rightarrow -1$ as $x \rightarrow -\infty$ and $w(x)
\rightarrow +1$ as $x \rightarrow +\infty$. In view of this, it turns out
convenient to consider the dependence of the independent variable $x$ as a
function of $w$, or rather the dependence of $u(w) \equiv
({{dx}\over{dw}})^{-1}$ as
a function of $w$. In fact, for a monotonic solution $w(x)$ of (1), $x(w)$
increases monotonically from $-\infty$ to $+\infty$ as $w$ goes from $-1$ to
$+1$. Thus, the function $u(w)$ is nonnegative and vanishes at both ends. Since
the original equation (1) is autonomous, one can rewrite it as a second order
equation for $u(w)$. In terms of $u$, $dw/dx = u$, $d^2w/dx^2 = u du/dw$
 and
$d^3w/dx^3 = {1\over 2} u d^2(u^2)/dw^2$. Therefore, equation (1) can be
rewritten as
\beq
{\frac{1}{2}} \lambda u {\frac{d^2 u^2}{dw^2}} + u = f(w), \quad w \in (-1,1)
\endq
together with the boundary condition $u(-1) = u(+1) = 0$. This is a nonlinear
second order differential equation for $u(w)$ singular at both endpoints.

In order to prove the desired bound on $\lambda$ multiply (2.1) by $g(w)/u$,
where $g(w)$ is any continuous function such  that $g(w)$ is twice
differentiable, $g(\pm 1)= 0$ and $g(w)$ is concave (i.e. $- g'' \ge 0$). A
specific choice for $g$ will be done shortly. Hence we have
\beq
{\frac{\lambda}{2}} g(w) {\frac{d^2 (u^2)}{dw^2}} + g(w) = {\frac{f(w)}{u}}
g(w).
\endq
We now integrate (2.2) in $w$ between $-1$ and $1$. After integrating by
parts the first term in the left side we  obtain
\beq
{\lambda\over 2} \int_{-1}^{1} g'' u^2 dw +\int_{-1}^{1} g(w)dw
  = \int_{-1}^{1}{{f(w)}\over u} g(w) .
\endq
Notice that when integrating by parts we have used that both $g$ and $u$ vanish
at the endpoints. Let $h = - g''$. Since $g$ is concave, $h$ is positive. From
(2.3) we have
\beq
\int_{-1}^{1} g(w) dw = \int_{-1}^{1}\left( {{f(w)}\over u} g(w) +
{\lambda\over
2} h u^2 \right) dw.
\endq
Since $f$, $g$ and $h$ are positive in $(-1,1)$ and $\lambda$ is a positive
constant, for any fixed $w$ we have that
\beq
\frac{f(w)g(w)}{u} + \frac{\lambda}{2} h(w) u^2 \ge {3\over 2} (fg)^{2/3}
\lambda^{1/3} h^{1/3},
\endq
(just minimize the right side as a function of $u$ for $u\in (0,+\infty)$).
{}From
(2.4) and (2.5) we have
\beq
\lambda^{1/3} \le {2\over 3} \frac{\int_{-1}^{1} g(w) dw}{\int_{-1}^{1}
(fg)^{2/3} h^{1/3} dw } .
\endq
The bound on $\lambda$ given by  (2.6) holds for any function $g$ twice
differentiable in $(-1,1)$ such that $h = - g''  \ge 0$ and $g(\pm 1) = 0$. If
$\lambda$ is larger than the right side of (2.6) for fixed $f$ and any such
$g$,
equation (1) cannot have monotonic fronts. For explicit examples of $f's$ one
can use directly (2.6) to derive upper bounds on $\lambda$. However, here we
would like to express a bound on $\lambda$ solely in terms of $f$ (i.e. an
explicit generic bound on $\lambda$). It is for this reason that we will pick a
specific $g$ in order to prove our main result. So choose $g$ in such a way
that
$h = - g'' = f$ in $(-1,1)$ and $g(-1) = g(1)$. Such a $g$ can be written
explicitly in terms of $f$ as
\beq
\int_{-1}^{1} K(s,t) f(t) dt
\endq
with $K(s,t) = {1\over 2} (s+1)(1-t)$ for $-1 \le s <t$ and
$ K(s,t) = {1\over 2} (1+t)(1-s)$ for $t < s \le 1$. With this particular
choice
of $g$, the bound (2.6) can be expressed as
\beq
\lambda^{1/3} \le {2\over 3} \frac{\int_{-1}^{1} g(w) dw}{\int_{-1}^{1}
f g^{2/3}  dw }  ={2\over 3} \frac{\int_{-1}^{1} g dw}{\int_{-1}^{1}
(- g'') g^{2/3}  dw },
\endq
and integrating by parts the denominator of the right side of (2.8) we get
\beq
\lambda^{1/3} \le  \frac{\int_{-1}^{1} g dw}{\int_{-1}^{1}
(g')^2 g^{-1/3}  dw }.
\endq
Writing $g = \psi^{6/5}$, the denominator $\int_{-1}^{1} (g')^2 g^{-1/3} dw$
becomes
$\frac{36}{25} \int \psi'^2$. Therefore
\beq
\lambda^{1/3} \le \frac{25}{36} \frac{\int_{-1}^{1} \psi^{6/5}
dw}{\int_{-1}^{1}
(\psi')^2 dw}.
\endq
Let $I$ denote the maximum of the quotient $ (\int_{-1}^{1} \psi^{6/5}
dw)^{5/3}/
\int_{-1}^{1} (\psi')^2 dw $ taken over all functions $\psi$ continuous on
$(-1,1)$ ( to be precise, the maximum of the quotient is taken over all
functions $\psi$ in the Sobolev space $H_0^1(-1,1)$ ). It is not difficult to
show
that this maximum does exist and that the corresponding maximizing function is
unique up to a multiplicative constant. In fact, the maximizing $\psi$
satisfies
the equation
\beq
-\psi'' = \psi^{1/5} \qquad {\rm in}\quad (-1,1)
\endq
together with the boundary conditions $\psi (-1) = \psi (1) = 0$. One can solve
numerically (2.11) and evaluate $I$. The numerical value of $I$ is
approximately
 .5549. From (2.10) we get
\beq
\lambda^{1/3} \le \frac{25}{36} \frac{I}{(\int_{-1}^{1} g dw )^{2/3}} ,
\endq
since $g = \psi^{6/5}$. Using (2.7) we can evaluate $\int_{-1}^{1} g dw$
explicitly in terms of $f$. We have $\int_{-1}^{1} g dw
= \int_{-1}^{1}\int_{-1}^{1} K(w,t) f(t) dt dw = \int_{-1}^{1} f(t)
\{ \int_{-1}^{t}K(w,t) dw + \int_{t}^{1}K(w,t) dw \} dt = {1\over 2}
\int_{-1}^{1} f(t) (1-t^2) dt $ so finally we get our bound
\beq
\lambda \le \frac{.228}{\left[ \int_{-1}^{1} f(t) (1-t^2) dt \right]^2} .
\endq
Hence if, for a given $f$, $\lambda$ is larger than the right side of (2.13),
equation (1) has no monotonic fronts.

\section{Applications}

\setcounter{equation}{0}
We first consider the equation for needle crystals including anisotropy. This
corresponds to our equation (1.1) with
 \beq
f(w) = \cos ({{\pi w}\over{2}})/(1 + \alpha \cos (2\pi w)),
 \qquad 0 < \alpha < 1 .
\endq
In this case it has been shown \cite{KS91}
 that monotonic fronts exist for a discrete set of
values of $\alpha$ and small $\lambda$. This $f$ vanishes at $w = \pm 1$ and,
for $0 < \alpha < 1$, f is positive so our theorem applies here.

If we insert $f(w)$ given by (3.1) in equation (2.13) we get an upper bound
$\lambda_u (\alpha)$ on the possible values of $\lambda$ for which one could
have monotonic fronts. This function $\lambda_u (\alpha)$ is shown in Figure 1.
Note that $\lambda_u (\alpha)$ is decreasing, $\lambda_u (0) = .214$ and
$\lambda_u (1) = 0$.

As a second example we consider
an exactly solvable model given by equation (1.1) with
\beq
f(w) = {1\over 2} (1- w^2) (1- {\alpha \over 2} + {3\over 2} \alpha w^2),
\qquad 0 < \alpha < 2.
\endq
In this case, monotonic fronts exist when $\lambda = \alpha$. In fact the
solution of equation (1.1) with $f$ given by (3.2) and $\lambda = \alpha$ is
given by $w(x) = \tanh ({x\over 2})$. The function $f$ given by (3.2) vanishes
at $w = \pm 1$, and for $0 < \alpha < 2$ it is positive, so again in this case
our theorem applies. Inserting (3.2) in (2.13) we get an explicit bound
 $\lambda_u (\alpha)$ given by
\[
\lambda_u (\alpha) = \frac{39.2765}{(7 - 2\alpha)^2}.
\]
In Figure 2 we have plotted this bound. The solid line corresponds to
$\lambda_u
(\alpha)$ while the dotted line corresponds to $\lambda = \alpha$, the exact
value for which there is a front.

As a final remark, we wish to point out that, if in a particular case a better
bound is sought, one may go back to equation (2.6) and find the best $g$ for
the problem. The method presented here can also be used in equations of the
form $\lambda w''' + w'' + w' = f(w)$ , with $\lambda > 0$, $f(\pm1) = 0$ and
$f$
positive and continuous between -1 and 1. In order to get a bound for this
equation an adequate choice for the trial function $g$ has to be made. The
choice depends on $f$. Some results
for $f(w) = 1- w^2$ are given in (\cite{BD93}).

\section{Acknowledgments}

We thank M. S. Ashbaugh for helpful suggestions. This work was partially
supported by Fondecyt project 1930559.

\newpage
\begin{centerline}
{\bf Figure Captions}
\end{centerline}
\vspace{1.0cm}

Figure 1.

Upper bound on the value of $\lambda$ for the existence of monotonic fronts in
the geometric model of crystal growth with anisotropy.

\vspace{.5cm}
Figure 2.

The solid line depicts the upper bound on the value of $\lambda$ for the
existence of fronts of the exactly solvable example. The dotted line
corresponds
to the values for which there is a solution.


\begin{thebibliography}{10}

\bibitem{BKKL84}
R. Brower, D. Kessler, H. Levine, and J. Koplik 1984
\newblock {\it Phys. Rev. A} {\bf 29} 1335

\bibitem{BGKL84}
E. Ben-Jacob, N.~D. Goldenfeld, G. Kotliar, and J.~S. Langer 1984
\newblock {\it Phys. Rev. Lett.} {\bf 53} 2110

\bibitem{KSuzuki76}
Y. Kuramoto and T. Tsuzuki 1976
\newblock {\it Prog. Theor. Phys.} {\bf 55} 356

\bibitem{Siv77}
G. Sivashinsky 1977
\newblock {\it Acta Astronautica} {\bf 4} 1117

\bibitem{Toland88}
J.~F. Toland 1988
\newblock {\it Proc. Roy. Soc. Edin.} {\bf 109A} 23

\bibitem{JTM92}
J. Jones, W.~C. Troy, and A.~D. MacGillivary 1992
\newblock {\it J. Diff. Eqns.} {\bf 96} 28

\bibitem{GH91}
R. Grimshaw and A.P. Hooper 1991
\newblock {\it Physica D} {\bf 50} 231

\bibitem{HM89}
J.~M. Hammersley and G. Mazzarino 1989
\newblock {\it IMA J. Appl. Math.} {\bf 42} 43

\bibitem{AM90}
C.~J. Amick and J.~B. McLeod 1990
\newblock {\it Arch. Rational Mech. Anal.} {\bf 109} 139

\bibitem{KS91}
M. Kruskal and H. Segur 1991
\newblock {\it Stud. Appl. Math.} {\bf 85} 129

\bibitem{Grimshaw92}
R. Grimshaw 1992
\newblock {\it Wave Motion} {\bf 15} 393

\bibitem{BD93}
R.~D. Benguria and M.~C. Depassier 1993
\newblock On the monotonic fronts of a dispersive {Kuramoto-Sivashinsky}
  equation
\newblock {\it preprint}
\end{thebibliography}
\end{document}